# TO PARALLELIZE OR NOT TO PARALLELIZE, BUGS ISSUE

A.  I.  El-Nashar             N.  Masaki

Computer Science Department, Faculty of Science,
Minia University, Minia - Egypt
*nashar_al@yahoo.com*

School of Electrical and Computer Engineering,
Kanazawa University -  Japan

***Abstract***: *Program correctness is one of the most difficult challenges in paralle  programming. Message Passing Interface MPI is widely used in writing parallel applications. Since MPI is not a compiled language, the programmer will be enfaced with several programming bugs.This paper presents the most common programming bugs arise in MPI programs to help the programmer to compromise between the advantage of parallelism and the extra effort needed to detect and fix such bugs. An algebraic specification of an MPI-like programming language, called Simple MPI (SMPI), to be used in writing MPI programs specification has also been proposed. In addition, both nondeterminacy and deadlocks arise in SMPI programs have been verified using Maud system.*

## 1.  Introduction

Debugging parallel programs is classified as a difficult problem. Most of parallel Application programmers focus only on the constructive part of creating a parallel algorithm for a particular problem and how to implement it, but ignore the issues of debugging [1]. Parallel programming adds a new category of bugs caused by the interaction of multiple communicated parallel tasks.  These parallel bugs 'Heisen bugs' [6], [14] are difficult to be detected and resolved due to the nondeterministic nature of parallel tasks running which makes the bugs may disappear when one attempt to detect them. The problem is that, for every bug, there is a reason, for every reason, there is a bug fix. The hardest step in solving this kind of bugs is working backward from a software failure to the original program error. The overhead of locating program bugs may make application programmers thinking a lot to decide: "to parallelize or not to parallelize," and as a result, they may prefer using the ordinary serial programming style to avoid extra bugs detecting and fix effort. There is a bad need of a reliable debugging tool that enables the user to detect such bugs since conventional debugging techniques fail in detecting all types parallel bugs. Message Passing Interface, MPI [25] is widely used for writing parallel programs. MPI library allows a user to portability compile his applications using C, C++, and FORTRAN, and then run on a selected number of processors. Since MPI itself is considered as a library not a compiled language; many programming bugs may occur. Most of these bugs can't be detected by the compiler of the used language. This paper presents the most common programming bugs arise in MPI programs as a type of parallel programs trying to achieve a bugs free MPI program. Also we propose an algebraic specification of an MPI-like programming language, called Simple MPI (SMPI), that can be used in writing MPI programs specification. Maud system [24]  has been employed to verify both nondeterminacy and deadlocks arise in SMPI programs.

The paper is organized as follows: section 2 presents the fundamentals of MPI programming. In section 3, the different MPI programming errors are presented. Section 4, concerns with MPI programs debugging techniques. In section 5, we describe an algebraic specification of an MPI-like programming language SMPI which supports message passing between several processes with functions "send" and "receive".

## 2. MPI Fundamentals

The goal of MPI is to establish a portable, efficient, and flexible standard for message passing to be used for writing message passing programs, and providing an appropriate environment for general purpose message-passing programs, especially programs with regular communication patterns. MPI is a library of subroutines that enables programmers to identify and implement explicit parallelism using special constructs. It provides all the subroutines that are needed to divide the tasks involved in the computational process into subtasks that can be distributed among a number of processors. MPI uses objects called communicators and groups to define which collection of processes may communicate with each other. A communicator must be specified as an argument for most MPI routines. The predefined communicator MPI_COMM_WORLD is used whenever a communicator is required, it includes all of MPI processes. Within a communicator, every process has its own unique, integer identifier "rank" assigned by the system when the process initializes. Ranks are contiguous and begin at zero, used by the programmer to specify the source and destination of messages, and also used conditionally by the application to control program execution. MPI contains more than one hundred functions that greatly ease the tasks in implementing common communication structures, such as send, receive, broadcasts and reductions. Several implementations exist to support a wide range of major hardware and software platforms. There are several MPI implementations. Each implementation provides a different mechanism for compiling MPI programs. This paper concerns with compiling and running MPI programs with MPICH2 [7], [27] under Windows Xp. First of all, the path of "mpich2\lib" and "mpich2\include" must be added to the system environment variables list to enable the compiler to access the required library and include files [26].

MPICH2 provides an installed C++ compiler, so the compilation and linking process is performed as in the ordinary case, in addition, the library file "mpi.lib" must be added to the "Object/library modules" . To compile an MPI FORTRAN source code, a FORTRAN compiler must be installed. The compilation mechanism is performed as in case of compiling the ordinary FORTRAN source codes, but the library files "fmpich2.lib", "fmpich2s.lib", and "fmpich2g.lib" must be added to the end of the "Object/library modules" to enable the linker to access the required object files during building the executable.

## 3. MPI Errors

MPI programs are coded in a special manner, in which each process executes the same program with unique data. All parallelism is explicit; the programmer is responsible for correctly identifying parallelism and implementing parallel algorithms using MPI constructs. MPI programming is an error prone process in all of its phases starting from coding, compiling, and linking ending with running. In this paper, we classify MPI errors into two main categories: compile-link time errors and run time errors. Each category demonstrates several error types.

### 3.1 Compile-Link Time Errors

An MPI program consists of four parts. Each part has its own "Compile-link time" error as described below.

a)      The first code part is the MPI include file which is required for all programs/routines which make MPI library calls. The user code must contain one of the header files "mpi.h" for C/C++, and "mpif.h" for Fortran. These files contain definitions of constants and prototypes required during compiling  source codes that contain MPI library calls. If the include file is not included in the source program, the used language compiler does not detect any errors in both compilation and linking phases even if the program contains MPI function calls. The effect of the missed include file appears only at run time. On other hand, the compiler can detect the pointer arguments errors in case of using MPI_Status and MPI_TAG in compilation phase.

b)      The second part is responsible for initializing MPI environment. MPI environment management routines are used for an initializing and terminating the MPI environment, querying the environment and identity. MPI is initiated by a call to MPI_Init. This MPI routine must be called before any other MPI routines and it must only be called once in the program. If this call is not included in the source program, called more than once, or  if there are MPI calls before it, the used language compiler does not detect any errors in both compilation and linking phases. The effect of the missed call appears only at run time.

c)      The third part is the body of program steps, calculations, and message passing calls. C/C++ MPI function calls start with "MPI_" followed by a character string with the leading character in upper case letter while the rest in lower case letters. Fortran subroutines bear the same names but are not case sensitive. The only difference between MPI subroutines (for Fortran programs) and MPI functions (for C/C++ programs) is the error reporting flag. In FORTRAN, it is returned as the last argument in the subroutine's argument list. In C/C++, the integer error flag is returned through the function value. The compiler can only detect the errors arising from calling MPI functions/subroutines with different number of actual arguments but fails in detecting a negative message length and also the errors arise from messages which exceed the bounds of the source/destination array .

d)       The last part is terminating MPI environment. MPI processing ends with a call to MPI_FINALIZE. The used language compiler does not detect any errors in both compilation and linking phases in the following cases: if the finalize call is not included in the source program and /or if there are MPI calls after MPI_FINALIZE.

### 3.2 Run Time Errors

To execute an MPI executable application, both the execution path and environment variables must be checked. MPICH2 provides a tool that initializes the parallel program execution called mpiexec. The user changes to the directory where his executable is located and issues the command  "mpiexec –np x fname",  where x is the number of processors and fname is the name of the executable file. When an MPI program starts, the program spawns into the number of processes as specified by the user. Each process runs and communicates with other instances of the program, possibly running on the same processor or different processors. The successful operation of building MPI executable does not guarantee that the application is bugs free. Several  MPI bugs can't be detected during compilation and link operations as shown in the previous section. Most of these bugs arise during application running.
Nondeterminacy and deadlock are two types of bugs that can't be detected during linking and compilation phases. In many cases, communications among running processes lead to either a nondeterminate results or deadlock.

Nondeterminate results arise according to the nondeterministic data arrival order to the process that computes the results and hence the output results may vary for multiple runs of the same program with the same input data. Deadlocks errors arise if the dependencies between parallel tasks described by an MPI construct can't be satisfied. Deadlocks are classified into "actual" and "potential". An "actual" deadlock will occur in the following cases:

a)  A process executes the MPI synchronous send, MPI_SSEND, and there is no corresponding call to an MPI receive routine.
b)  A process executes a receive routine and there is no corresponding call to a send routine

A "potential" deadlock occurs when a process executes the MPI standard send, MPI_SEND, without corresponding MPI receive, in this case the call to MPI_SEND copies the message to a buffer (only if MPI implementation provides send buffer) and execution continues. If the message was not copied the message to the buffer, a deadlock would have occurred.
Verification of both nondeterminacy and deadlock will be studied in section 5.

## 4. MPI Programs Debugging Techniques

Many techniques are used for locating parallel programming bugs. We give a brief discussion of the most commonly used techniques, namely dynamic analysis, static analysis and model-based test.

### 4.1 Dynamic analysis

Dynamic analysis implies the necessity of launching an application and executing different sequences of operations to analyze the program behavior. There are four popular parallel debugging techniques: printf-style debugging, launching serial debuggers in parallel, attaching serial debuggers to individual parallel processes, using parallel debuggers.

### 4.1.1 Printf-style debugging

Inserting a "Print" statement prior to or after a specific point in the source program to print out, for example, the value of a program variable at that point in the program to a file or a standard output is one of the most common forms of debugging. This debugging technique is not a powerful one in case of MPI programs because of the following reasons:

a)      The application must be edited, re-compiled, and re-run, each time the programmer needs to test other program statement(s),
b)      The standard output from the print statement will make some delay before its displaying on the standard output which may negatively affects the application's speed up,
c)      The output messages may be displayed in an interleaved manner that does not reflect the actual processes execution order, or it may be buffered by the run-time system or MPI implementation, so it will not be displayed at all even the print statement was executed.

Intel Message Checker (IMC) [11] extends this approach to perform a post-mortem analysis by collecting all information on MPI calls in a trace file. After program execution, this trace file is analyzed by a separate tool or compared with the results from previous runs [14].

### 4.1.2 Launching serial debuggers in parallel

The traditional serial debuggers can be used to debug MPI applications by setting breakpoints to investigate a specific state. The Debugger allows the programmer to single-step through his running application to test a process against a specific fault. GNU debugger (gdb) [21] and its GUI counterpart like Data Display Debugger (ddd) [2] can be used to debug MPI parallel applications. In practice, the programmer should issue the "mpiexec" command with the appropriate switches to launch the MPI processes under the serial debugger. For example, the command "mpiexec -np 4 -e gdb mycode" will launch four copies of gdb and load the executable "mycode". The user can then individually control each MPI process.

Launching serial debuggers in parallel is useful but has several drawbacks:

a)   It is suitable only for small and mid-sized parallel runs
b)   Inserting breakpoints during debug operation affects processes timing leading to extra timing-related bug.
c)   Stopping a single process for a debugging reason and making the rest of the parallel processes continue to run will cause several problems if there is a dependency between the  running processes and the stopped one.
d)   Once the processes run completes, the programmer has to exit the debugger and re-run the MPI application for another debug session.

### 4.1.3 Attaching serial debuggers to individual parallel processes

In this approach, the programmer can use "attach to processes" feature of a serial debugger to attach serial debugger to one or more specific running MPI process(es) not to all the MPI application processes. In contrast to that method  described in the previous section, this method is more portable since it does not depend on the capabilities of the MPI implementation. All of programmer's need is to know the process id to which he wishes to attach the debugger. Practically, this can be achieved by inserting an infinite loop containing the statement "getpid()" followed by "sleep()" in the considered application at the point the programmer wants to attach the debugger. The "getpid()"  statement  will display the required PID to be attached, the sleep() function will enforce the required process to  wait forever enabling the programmer to attach with a debugger.  This method  is not a powerful one but is considered very  practical.

### 4.1.4 Using parallel debuggers

Another approach to debug MPI applications is to develop dedicated debugging tools "parallel debuggers", to find MPI applications bugs at runtime. These tools detect violations of programming rules imposed by MPI  such as bounds of the message buffer.
There are different message-checking tools that use such idea like MPI-CHECK [10], Umpire [17] and MARMOT [3],[4]. These debuggers are effective in detecting some types of software bugs at runtime but still poor to detect semantics-related bugs [18], [19], and also highly dependent on domain-specific expertise and human efforts [9],[15].

### Static analysis

Static analysis approach handles only the source code without its execution. This approach can be useful to determine detailed and full coverage of the analyzed code. In case of MPI programs, static analysis can detect errors that may not appear during real program execution, and hence it can complement dynamic analysis to discover more bugs. On other hand, static analysis approach suffers from some drawbacks:

a)    Static analyzer is very complicated to be implemented.

b)    A huge space of possible states appears and the size of the tested information becomes unacceptably large.

c)    It may display a lot of false messages about potential errors and demand great effort to minimize their number.

d)    It requires an intermediate source code representation such control flow graphs as in case of data flow testing. This means that extra effort has to be done in building CFG representing MPI programs MPI-CFG [8]  since the ordinary CFG does not demonstrate most of MPI constructs like inter-process communication and synchronization edges.

### Model-based

Model-based testing is a software testing approach in which test cases are derived from a model that describes the system under test. Practically, model-based test works only for small base blocks of an application. In most cases it is very difficult to automatically build a model on the basis of the code, also the manual creation of models is a hard and error prone process. This approach suffers from the problem of quick extension of state space that is can be partially controlled by using reduction methods. For MPI programs, this approach would require that programmers build, either manually or automatically, a model of their applications in a language such as MPI-SPIN [22],[28], or Zing [20]. The MPI program model creation adds an extra overhead to the programmers' tasks.

## 5. MPI Verification

In this section, we describe an algebraic specification of an MPI-like programming language, called Simple MPI (SMPI), which supports message passing between several processes with functions "send" and "receive". The specification of SMPI is described in the algebraic specification language Maude [16],[24], which is a member of the OBJ family, and is a successor of the algebraic specification language OBJ3 [13]. As related studies, Salman et al have shown how formal verification based on model checking can be used to find actual deadlocks in algorithms that use the MPI one-sided communication primitives [23]. Goguen and Malcolm have proposed algebraic semantics of imperative programs in the algebraic specification language OBJ3 [12].  Nakamura et al. have proposed a behavioral specification of imperative programs in CafeOBJ [17]. CafeOBJ [5] is another successor of OBJ3. In this section, we give a rewrite specification of parallel imperative programs in Maude. One of the strong points of rewrite specifications is that it provides automatic exhaustive searching for verifying a given specification.

### 5.1 Syntax of *SMPI*

SMPI programs are coded in a manner that is very close to ordinary C++ and FORTRAN MPI programs which makes SMPI representation of the ordinary MPI codes is very simple. SMPI deals with the fundamental programming constructs such as variables, types and expressions. In addition to these fundamentals, we introduce a simplified version of send function, which corresponds to the synchronous send function MPI_SSEND and also two versions of recv function that corresponds to the ordinary MPI_RECV of  MPI, as functions for message-passing in SMPI. The syntax of these constructs are described as follows:

a)    All variables must be expressed as lower-case letters a, b, c, …, z with two special variables, pid and  np, which represent process ID and the number of processes respectively.

b)    The primitive type assumed is only integers.

c)    Expressions in are constructed from variables, integers and operators +, *, > , etc.,. SMPI can deal with expressions like $(10 + x) * y > 0$. It also deals with fundamental program statements such as variables declaration, assignments, conditionals, and iterations.

d)    The syntax of SMPI synchronous send is **send (*message, destination*)**, where ***message*** is the data item to be sent from the process that includes this function call to the process whose ID is ***destination***. This function corresponds to the ordinary C++ MPI synchronous send which its syntax is: **MPI_SSend (&buf , count, datatype, dest, tag, comm)**.

e)    The syntax of the first version of receive function is **recv (*message, source*)** , where ***message*** is the data item to be received at the process that includes this function call from the process whose ID is ***source***. This function corresponds to the ordinary C++ MPI receive which its syntax is: **MPI_Recv (&buf,count,datatype,source,tag,comm,&status)**.

f)    The second version syntax is **recv (*message*,any)**, where ***message*** is the data item to be received at the process includes this function call from **any** other process. This function corresponds to the ordinary C++ MPI receive which its syntax is : **MPI_Recv (&buf, count, datatype, MPI_ANY_SOURCE, MPI_ANY_TAG, comm, &status)**. A typical example of our *SMPI* programs that uses the previous constructs is listed in Fig 5.1.

```
 1.  if(not(pid = 0)){
 2.     send(pid,0);
 3.  }
 4.  if(pid = 0){
 5.     int x ; int y ; int i ;
 6.     y := 1 ; i := np ;
 7.     while (i > 1) {
 8.        i := i - 1 ;
 9.        recv(x,i) ;
10.        y := y * x ;
11.     }
12.  }
```

Figure. 5.1: *SMPI* program P1

The program consists of two parts by the case splitting with respect to pid: the first part is represented by program lines 1:3 , and the latter one is represented by the lines 4:12 . In the first part, all processes except process 0 execute the body of the first conditional statement. The body part consists of only send(pid, 0), which means that each process will send its ID to process 0. In the later part, only the process 0 executes the body of the second conditional statement. In its body part, (line 5 and line 6), variables x, y and i are declared, 1 is assigned to y and the number of processes (np) is assigned to i.

The main part , (line 7 : line 11), is a while-loop whose condition is i > 1. The while-loop, repeatedly calls recv(x,i) from i = np to i= 1 after reducing the counter i by 1 to accommodate with the fact that the running processes IDs start with 0 and ends with np-1. After each call of recv, the variable y is multiplied by the received variable x.

In this example, The process 0 repeatedly receives the messages from all other processes whose IDs are np-1, np-2, …, 2, 1. Therefore, even if a process P sends a message before another process Q whose ID is larger than P's ID, the process P should wait until the process 0 receives all messages sent from the processes whose IDs are larger than P's ID. We predict that the final result will be the same for all multiple runs with the same number of processes. Thus, if the program is executed with n processes, then the value of y of the process 0 should be (n-1)!, for example, if n = 5, then the value of y will be 24.

### 5.2 Semantics of *SMPI*

We first give semantics of program execution with a single process. A snapshot of program execution can be formalized as a table of the declared variables and their values. We call it a store. A program can be regarded as a function on the set of stores. For example, the program x := y + z ; takes the store (x :: 1) (y :: 2) (z :: 10) and returns the store (x :: 12) (y :: 2) (z :: 10). For a store S and a program P, the term S P stands for the result of applying the program P to the store S. The semantics of execution of a sequence of programs P0 P1 P2… Pn is given as a rewrite sequence on terms as follows: S0 (P0 P1 P2… Pn ) => S1 (P1 P2… Pn ) => S2 (P2 … Pn ) =>…=> Sn Pn => Sn+1, where S0 is the empty initial store, denoted by init, and each store Si+1 is obtained by applying Pi to Si. For example, the execution of the sample program int i ; int x ; i := 1 ; x := i + 1 ; is given as follows (Fig. 5.2):

```
    init (int i ; int x ; i := 1 ; x := i + 1 ;)
=> (i :: na) (int x ; i := 1 ; x := i + 1 ;)
=> (i :: na)(x :: na) (i := 1 ; x := i + 1 ;)
=> (i :: 1)(x :: na)  (x := i + 1 ;)
=> (i :: 1)(x :: 2)
```

Figure. 5.2:  Execution of an *SMPI* sample program

Here, na stands for "not available", and (x :: na) means that no value is assigned to the variable x. A snapshot of parallel execution is given as a list S0 | S1 |S2 | … |Sn of stores  where each Si corresponds to the processor i. The semantics of parallel execution of a program P with n processors is specified as follows:  mpirun(n,P) => (pid :: 0) (np :: n)  P |(pid :: 1) (np :: n)  P |(pid :: 2) (np :: n)  P |…|(pid :: n) (np :: n)  P => …, where np is a variable which the number of processes is assigned to.

For each process i, the initial state is given as (pid :: i) (np :: n)  P and after the initialization, each process's state is rewritten in parallel.

Semantics of send and receive messages is given as the following conditional rewrite rule (Fig. 5.3):

```
crl ((S1 send(X1, Dest);) P1) | ((S2 recv(X2, Source);) P2)
 =>  (S1 P1)                   | (update(X2, S1[X1], S2) P2)
if S1[Dest] == S2[pid] and S1[pid] == S2[Source] .
```

Figure 5.3 Semantics of message passing in P1

A rewrite rule in Maude specifications describes a local concurrent transition. If there is an instance of the left-hand side (between crl and =>) of the rewrite rule and its condition part (between if and the period) is satisfied, then the state is rewritten into the state where the instance of the left-hand side is replaced with the corresponding instance of the right-hand side (between => and if). Thus, the above rewrite rule describes the following cases :

a)      If one process P is trying to execute send(X1,Dest) which sends a message X1 to the process Dest,

b)      If another process Q is trying to execute recv(X2,Source) which receives a message from the process Dest1 and assigns the message to the variable X2, and

c)      If the destination of the send message is the process Q and the source of the receive message is the process P, then the send and receive functions are consumed and the value of the variable X2 of the process Q is updated as the value of the variable X1 of the process P.

To show the difference between **recv (*message,source*)** and **recv (*message,*any)**, consider another SMPI program P2 which exactly resembles the code of P1, listed in fig 1, except line 9 of P1 is replaced by the function call recv(x,any). In case of P2, a message sent from any process can be received by `recv(x, any)`. Thus, in the above situation, the message sent from the process P is received before the process Q, and the process P does not waste a time. We also predict that the final result will be the same for all multiple runs with the same number of processes. The conditional rewrite rule that expresses the semantics of receive function **recv (*message,*any)** is shown in (Fig.5.4). The only difference between this rewrite rule and that described in (Fig 5.3) is that the received message does not depend on the ID of the sending process, and hence the message that arrives first will be considered.

```
crl ((S1 send(X1, Dest);) P1) | ((S2 recv(X2, any);) P2)
 =>   (S1 P1)                  | (update(X2, S1[X1], S2) P2)
if S1[Dest] == S2[pid] .
```

Figure 5.4: Semantics of message passing in P2

### 5.3 Verification of *MPI*

One of the important features of the algebraic specification language is that specifications are executable. Maude system enables the programmer to apply the rewrite rules to a given specification repeatedly by issuing the Maud command "rewrite *t* .", where t is the considered specification. Maud system, then returns the result term in which the rewrite rules cannot be applied to anymore. Such a result term is called a normal form. The following is the execution result of rewriting the term mpirun(5, *P2*), where *P2* is the modified version of P1 that uses the function recv(X2, any).

We omit the program listing and also a part of the execution result by the dots (…) and modify line breaks to improve the appearance. Maude> is the command prompt of the Maude system. The Maude rewrite command is applied to the term from the second line to the sixteenth line. A normal form of the input term in the above example is shown in the last three lines. Notice that the result store of the process 0 can be seen at the last line (i :: 1)(x:: 4)(y:: 24)(pid:: 0)(np:: 5), where the value of y is 24 (= 4!) as we expected. In our specification, without any translation, an SMPI program itself can be treated as (a part of) a target term. At the fifth line from the bottom, we can see how many rewrite steps are included in the rewrite sequence to obtain the normal form. There are 798 rewrite steps between the input term and the normal form

```
Maude> rewrite
mpirun(5,…) .
…
rewrites: 798 in 0ms cpu (1ms real) (2046153 rewrites/second)
result List:
   (pid :: 1) np :: 5 | (pid :: 2) np :: 5
| (pid :: 3) np :: 5 | (pid :: 4) np :: 5
| (i :: 1) (x :: 4) (y :: 24) (pid :: 0) np :: 5
```

Figure 5.5: Parallel execution of P2

.

### 5.3.1 Verification of Nondetrminacy

We note that the above execution result is just one of the possible normal forms of the parallel execution. Thus, it does *not* guarantee that the value of `y` is 24 in *all* possible parallel executions. To verify such a property, the Maude `search` command is useful. When we input "`search` *term* `=>!` *pettern* `such that` *condition*" to the Maude system, all possible normal forms of *term* are searched to be checked whether each of them can be an instance of *pattern* and the instance satisfies *condition* or not. If such a normal form exists, the Maude system shows evidence to us. The following is an example of the Maude `search` command:

```
Maude> rewrite
mpirun(5,…)
=>! ((pid :: 0)(y :: Y:Int) S:Store | L:List)
such that (Y:Int =/= 24) .
…
No solution.
states: 38850  rewrites: 1327185 in 2995ms cpu (3049ms real)
(443070 rewrites/second)
```

Figure 5.6:  Verification of P2

In Fig. 5.6, the input term is the same with that of Fig. 5.5, and the pattern and the condition mean that the value of the variable `y` in the process 0 is not 24. Then, the Maude system returns the message "No solution", which means that there is no such normal form, that is, it guarantees that in all possible parallel executions, the value of `y` in the process 0 is 24. To verify the property, the Maude system checks 38850 states with 1327185 rewrite steps. Modifying the input program, P2, by changing "`y := y * x ;`" at the third line of while-loop into "`y := x - y ;`". Then the Maude `rewrite` command returns the result as follows:

```
result List:
   (pid :: 1) np :: 5 | (pid :: 2) np :: 5
| (pid :: 3) np :: 5 | (pid :: 4) np :: 5
| (i :: 1) (x :: 4) (y :: 3) (pid :: 0) np :: 5
```

Figure  5.7 Parallel execution of the modified P2

In Fig. 5.7, the value of the variable `y` is 3 = 4 − (3 − (2 − (1 − 1))). We check whether in all normal forms the value is also 3 or not by the Maude `search` command as follows:

In Fig. 5.8, ten solutions are returned. In Solution 1, the value of the variable `y` is 1. In Solution 2, it is 5, and in Solution 10, it is −3. The obtained solutions reflect the nondeterministic feature of MPI programs. The reason of nondeterminacy in this case is that using of recv (x, any) implies that the computation involved will be affected by the nondeterministic arrival order of x, and hence the final assigned value may varies for each run. This situation did not appear in case of using the expression " Y= Y*X " because multiplication is commutative and will not be affected by the arrival order of X, in contrast to the case of using the expression " Y=X-Y ", in which subtraction operation will be affected yielding these nondeterministic results. Thus, we conclude that this program is nondeterminate.

```
Maude> search mpirun(5,…) =>! ((pid :: 0) (y :: Y:Int) S:Store |
L:List) such that (Y:Int =/= 3) .
…
Solution 1 (state 67255) …
L:List --> (pid :: 1) np :: 5 | (pid :: 2) np :: 5
        | (pid :: 3) np :: 5 | (pid :: 4) np :: 5
S:Store --> (i :: 1) (x :: 3) np :: 5
Y:Int --> 1

Solution 2 (state 67256) …
L:List --> (pid :: 1) np :: 5 | (pid :: 2) np :: 5
        | (pid :: 3) np :: 5 | (pid :: 4) np :: 5
S:Store --> (i :: 1) (x :: 4) np :: 5
Y:Int --> 5

…

Solution 10 (state 67265) …
L:List --> (pid :: 1) np :: 5 | (pid :: 2) np :: 5
        | (pid :: 3) np :: 5 | (pid :: 4) np :: 5
S:Store --> (i :: 1) (x :: 1) np :: 5
Y:Int --> -3

No more solutions.
states: 67266  rewrites: 2176921 in 5349ms cpu (5467ms real)
(406918 rewrites/second)
```

Figure 5.8: Verification of an *SMPI* program (2)

```
Maude> rewrite mpirun(5,
  if(not(pid = 0)){
    int x ;
    send(pid,0);
    recv(x,0);
  }
  if(pid = 0){
    int x ; int i ;
    i := 1 ;
      while (np > i) {
        recv(x,any) ;
        send(x,i) ;
        i := i + 1 ;
      }
  }
) .
…
rewrites: 737 in 0ms cpu (3ms real) (930555 rewrites/second)
result List:
  (x :: 1) (pid :: 1) np :: 5 | (x :: 2) (pid :: 2) np :: 5
| (x :: 3) (pid :: 3) np :: 5 | (x :: 4) (pid :: 4) np :: 5
| (i :: 5) (x :: 4) ( pid :: 0) np :: 5
```

Figure 5.9: Parallel execution of P3

### 5.3.2 Verification of Deadlock

Besides the nondeterminacy check, our specification can be used to detect other kinds of bugs. One of the typical bugs in message-passing parallel programs is a deadlock. Consider the following example:

In the program P3 shown in Fig. 5.9, each processes except 0 tries to send its ID to the process 0, and then tries to receive a message from the process 0 and assign the message to the variable x. The process 0 tries to receive a message from any source and send it to each process in the ascending order. In the store of each process in the above result, we can see that its ID is assigned to x. However, as we mentioned, it does not guarantee that all possible parallel execution work well like that. Now, we check whether the process 0 finishes the program in all possible parallel execution or not. In Fig. 5.10, the pattern of the `search` command means that the process 0 stops running with some part of the program `P1 P2` remaining. The Maude system returns the six normal forms (solutions) which are matched with the pattern. In Solution 6, the process 0 stops with `P1:Pgm --> send(x,i);` as the head of the remaining program and the store `S:Store --> (i :: 3)(x :: 4)(np :: 5)`, which tells us that the process 0 stops when trying to send a message to process 3. We can see that the current state of the process 3 in the list of states (**L**) which contains (`pid :: 3`), and that the process 3 also stops with `send(pid,0);`. Both the process 0 and 3 try to send a message to each other, and fails into a deadlock. Since we treat *SMPI* programs themselves as terms to be rewritten, the search result is easy to be read. We can directly see the point of the problem in the program. The readability of not only specifications but also the results of their executions and the traces of their verifications is one of the most important features of algebraic specification languages.

```
Maude> search mpirun(5,…
) =>! (((((pid :: 0) S:Store) P1:Pgm) (P2:Pgm)) | L:List) .
…
Solution 1 (state 11078)
…
…

Solution 6 (state 16639)
states: 16860  rewrites: 664229 in 1504ms cpu (1701ms real)
(441482 rewrites/second)
L:List -->
    (x :: 1) (pid :: 1) np :: 5
| (x :: 2) (pid :: 2) np :: 5
| ((x :: na) (pid :: 3) np :: 5) send(pid,0); recv(x,0);
      if pid = 0{int x ; int i ; i := 1 ;
      while np > i{recv(x,any); send(x,i); i := i + 1 ;}} end
| ((x :: na) (pid :: 4) np :: 5) recv(x,0); if pid = 0
      {int x ; int i ; i := 1 ; while np > i{recv(x,any);
      send(x,i); i := i + 1 ;}} end
S:Store --> (i :: 3) (x :: 4) np :: 5
P1:Pgm --> send(x,i);
P2:Pgm --> i := i + 1 ; while np > i{recv(x,any); send(x,i);
          i := i + 1 ;} end

No more solutions.
states: 19232  rewrites: 729193 in 1679ms cpu (1878ms real)
(434063 rewrites/second)5
```

Figure 5.10: Verification of P3

## 6. Conclusion

Parallel programs are difficult to debug due to their nondeterministic features. There may be potential bugs lurking behind the cover of nondeterminism. In this paper, we have categorized the most common parallel programming bugs arise in MPI programs as a type of parallel programs. We have also summarized MPI programming debugging techniques and tools such as MARMOT, UMPIRE, IMC, and MPI-CHECK. We found that these tools are capable of detecting many errors in MPI programs but do not guarantee to explore systematically all the execution interleaving of a program. Also we deduced that no single method is superior and a variety of approaches need to be supported to achieve a reliable debugging features. Finally, we proposed an algebraic specification of an MPI-like programming language, called Simple *MPI* (*SMPI*). *SMPI* simplifies the programmer's task to write the corresponding specifications of ordinary C++ or FORTRAN MPI programs. The proposed specification deals with the fundamental programming constructs such as variables, types and expressions. The specification handles a simplified version of send function, which corresponds to the synchronous send function MPI_SSEND and also two versions of recv function that corresponds to the ordinary MPI_RECV of MPI. Nontederminacy and deadlocks in *SMPI* programs have been successfully verified by using Maud system .

## References

1.  A. Grama, A. Gupta, G. Karypis, and V. Kumar, "Introduction to Parallel Computing", 2nd edition, Addison Wesley, 2003.
2.  A. Zeller, "Debugging with DDD", User's Guide and Reference Manual, Version 3.2, Copyrightc 2000 Universität Passau, Lehrstuhl für Software-Systeme, Innstraße 33, D-94032 Passau, Germany, Last updated 2000-01-03
3.  B. Krammer, K. Bidmon, M. S. Mller, and M. M. Resch, "MARMOT: An MPI analysis and checking tool",In Parallel Computing (PARCO), Center for High Performance Computing Dresden University of Technology, Germany, 2003.
4.  B. Krammer, M. S. Muller and M. M. Resch, "MPI I/O Analysis and Error Detection with MARMOT", In Recent Advances In Parallel Virtual Machine And Message Passing,11th European PVM/MPI Users' Group Meeting. LNCS 3241, pp 242 - 250, Springer, 2004.
5.  CafeOBJ official homepage: http://www.ldl.jaist.ac.jp/cafeobj/index.html.
6.  E. McDowell and D. P. Helmbold, "Debugging Concurrent Programs", ACM Computing Surveys, December 1989.
7.  D. Ashton and J. Krishna, "MPICH2 Windows Development Guide", version 1.0.7, Mathematics and Computer Science Division Argonne National Laboratory April 2, 2008.
8.  D. Shires, L. Pollock, and S. Sprenkle, "Program flow graph construction for static analysis of mpi programs", In International Conference on Parallel and Distributed Processing Techniques and pplications, (PDPTA 99), June 1999.
9.  G. Luecke, H. Chen, J. Coyle, J. Hoekstra, M. Kraeva, and Y. Zou, " MPI-CHECK: a tool for checking Fortran 90 MPI programs", Concurrency and Computation: Practice and Experience, Volume 15, pp 93-100, 2003.
10. G. Luecke, Y. Zou, J. Coyle, J. Hoekstra and M. Kraeva," Deadlock Detection In MPI Programs", In Concurrency and Computation: Practice and Experience, Volume 14, pp 911 – 932, 2002.
11. J. DeSouza, B. Kuhn, and B. R. de Supinski, "Automated, scalable debugging of MPI programs with Intel message checker", In Proceedings of the 2nd international workshop on Software engineering for high performance computing system applications, Volume 4, pp 78–82, ACM Press, NY, USA, 2005.

12. J. Goguen and G. Malcolm, "Algebraic Semantics of Imperative Programs", MIT Press, 1996.
13. J. Goguen, T. Winkler, J. Meseguer, K. Futatsugi, and J. Jouannaud, "Introducing OBJ, In Software Engineering with OBJ" , Kluwer Academic Publishers, pp.3 -167, 2000.
14. J. Huselius," Debugging Parallel Systems: A State of the Art Report", MRTC Report no. 63, September 2002.
15. J. S. Vetter and B. R. de Supinski, "Dynamic software testing of MPI applications with Umpire", In ACM/IEEE Conf. on Supercomputing (SC), 2000.
16. Masaki Nakamura, Alaa Ismail El-Nashar, and Kokichi Futatsugi, "An algebraic specification of message passing programming languages", Forum on Information Technology FIT2009, Japan, pp. 85-92, September 2009.
17. M. Nakamura, M. Watanabe, and K. Futatsugi, "A Behavioral Specification of Imperative Programming Languages", IEICE Transactions on Fundamentals of Electronics, Communications and Computer Sciences, Volume E89-A, Number 6, pp 1558-1565, 2006.
18. N. Nethercote and J. Seward "Valgrind: A framework for heavyweight dynamic binary instrumentation", In ACM SIGPLAN Conference on Programming Language Design and Implementation (PLDI), San Diego, California, USA, 2007.
19. R. Hastings and B. Joyce, "Purify: Fast detection of memory leaks and access errors", In Winter USENIX Conference, 1992.
20. R. Palmer, S. Barrus, Y. Yang, G. Gopalakrishnan, and R. M. Kirby, "Gauss: A framework for verifying scientific computing software", In Workshop on Software Model Checking, 2005.
21. R. Stallman, R. Pesch, S. Shebs, et al, "Debugging with GDB, The gnu Source-Level Debugger", 9th Edition, for GDB version 5.2.1, Boston, MA 02111-1307 USA, ISBN 1-882114-77-9, December 2001.
22. S. F. Siegel, "Model checking nonblocking MPI programs", In Verification, Model Checking, and Abstract Interpretation (VMCAI), January 2007.
23. S. Pervez, G. Gopalakrishnan, R. M. Kirby, R. Thakur, and W. Gropp, "Formal verification of programs that use MPI one-sided communication" In EuroPVM/MPI, pp 30–39, 2006.
24. The Maude system: http://maude.cs.uiuc.edu/.
25. The MPI Forum. The MPI-2: Extensions to the Message Passing Interface, July 1997. http://www.mpi-forum.org/docs/mpi-20-html/mpi2-report.html.
26. W. Gropp and E. Lusk: MPICH2. http://www.mcs.anl.gov/mpi/mpich2/ ,2006 .
27. W. Gropp, E. Lusk, N. Doss, and A. Skjellum, "A high-performance, portable implementation of the MPI message  passing interface standard",  Parallel Computing, Volume 22, Number 6, pp 789–828, 1996.
28. Y. Yang, X. Chen, G. Gopalakrishnan, and R. M. Kirby, "Distributed dynamic partial order reduction based verification of threaded software", In Workshop on Model Checking Software (SPIN 2007), July 2007.